\documentstyle[multicol,aps,prl,floats]{revtex}

\begin{document}
\title{Intermittent dissipation of a passive scalar in turbulence}
\author{M. Chertkov$^a$, G. Falkovich$^b$ and I.Kolokolov$^c$}
\address{$^{a}$ Physics Department, Princeton University, Princeton, NJ 08544, USA\\
$^{b}$ Physics of Complex Systems, Weizmann Institute, Rehovot,
76100, Israel\\
$^{c}$ Budker Institute of Nuclear Physics, Novosibirsk 630090, Russia}
\date{September 2, 1997 }
\maketitle

\begin{abstract}
Probability density function (PDF) of passive scalar dissipation ${\cal P}%
(\epsilon)$ is found analytically in the limit of large Peclet and Prandtl
numbers (Batchelor-Kraichnan regime) in two dimensions. The tail of PDF at $%
\epsilon\gg\langle\epsilon\rangle$ is shown to be stretched exponent $\ln%
{\cal P}(\epsilon)\propto\epsilon^{1/3}$. At $\epsilon\ll\langle\epsilon%
\rangle$, ${\cal P}\propto 1/\sqrt{\epsilon}$.
\end{abstract}

\pacs{PACS numbers 47.10.+g, 47.27.-i, 05.40.+j}
\draft

\begin{multicols}{2}

Probability distribution of the gradients of turbulent fields is probably
the most remarkable manifestation of the intermittency of developed
turbulence and related strong non-Gaussianity. A typical plot of the
logarithm of gradient's PDF (which would be parabolic for Gaussian
statistics) is concave rather than convex, with a strong central peak and
slowly decaying tails. This is natural for an intermittent field since rare
strong fluctuations are responsible for the tails while large quiet regions
are related to the central peak. In particular, such PDFs were observed for
the dissipation field (square gradient) of passive scalar advected by
incompressible turbulence which is the subject of the present paper. We
consider scalar advection within the framework of the Kraichnan model
assuming velocity field to be delta-correlated in time \cite{74Kra-a}. Most
of the rigorous results on turbulent mixing has been obtained so far with
the help of that model which is likely to play in turbulence the role Ising
model played in critical phenomena. High-order moments of the scalar were
treated hitherto by the perturbation theory around Gaussian limits. Clearly,
the kind of strongly non-Gaussian PDF observed for gradients cannot be
treated by any perturbation theory that starts from a Gaussian statistics as
zero approximation.

Since we consider developed turbulence with large Peclet number $Pe$
(measuring relative strength of advection with respect to diffusion at the
pumping scale), it is tempting to use $Pe^{-1}$ as a small parameter. Yet
any attempt to treat diffusion term perturbatively is doomed to fail because
the PDF of the dissipation is a nonperturbative object with respect to the
inverse Peclet number: it is zero at $\epsilon\neq 0$ and zero diffusivity
yet have nonzero limits as diffusivity goes to $+0$. Indeed, the main
contribution into dissipation is given by the scales around the diffusion
scale where advection and diffusion are comparable. Still, the presence of a
small parameter calls for finding a proper way to simplify the description.
Following \cite{74Kra-a,94SS,95CFKLa,97CKVa}, we show here that the
dynamical formalism of Lagrangian trajectories is a proper language to
describe the probability distribution of the dissipation field. Our
goal is to express unknown (statistics of dissipation) via known (statistics
of pumping). Since Peclet number is the ratio between pumping and diffusion
scales, then any piece of scalar has a long way to go between birth and
death, our goal is to describe how statistics is modified along the way.
Dynamical formalism explicitly reveals the presence of two different time
scales, a short one related to stretching and a long one related to
diffusion (which eventually restricts the process of stretching); this time
separation has been exploited first in solving similar problem for
one-dimensional compressible velocity \cite{97CKVa}. Time scale of
stretching fluctuations is of order of inverse Lyapunov exponent while the
whole time of stretching is $\ln Pe$ times larger. Clear time separation in
the dynamical history is guaranteed for (but probably not restricted by)
trajectories that correspond to the value of $\epsilon$ larger than that
produced by the pumping, $\langle\epsilon\rangle/Pe^2$, yet smaller than $%
\langle\epsilon\rangle\ln Pe$ (whether it is possible to extend the
time-separation procedure for a wider interval of $\epsilon$ will be the
subject of future analysis \cite{97CFKb}). At $Pe\to\infty$, we are able to
calculate PDF ${\cal P}(\epsilon)$ rigorously, exploiting time separation
and executing explicitly separate averaging over slow and fast degrees of
freedom. 

There are three main steps in deriving ${\cal P}(\epsilon)$
:\newline
{\bf 1}. Presenting the PDF as an average of a functional of the
time-ordered exponent of the strain matrix;\newline
{\bf 2}. Reparametrization of the problem into nonlocal auxiliary quantum
mechanics 
\newline
{\bf 3}. Implementing time separation which makes reduction to two coupled
yet local quantum mechanics: one-dimensional one describing the long
evolution 
and two-dimensional one describing the fast fluctuations.

Let us define the problem and make the first step of the derivation.
Advection of a passive scalar $\theta (t,{\bf r})$ by an incompressible flow
${\bf v}(t,{\bf r})$ is governed by the equation
\begin{equation}
(\partial _t+v_\alpha \nabla _\alpha -\kappa \triangle )\theta =\phi
\,,\quad \nabla _\alpha v_\alpha =0\,,  \label{a1}
\end{equation}
where $\phi (t,{\bf r})$ is the external source and $\kappa $ is the
diffusivity. Both ${\bf v}(t,{\bf r})$ and $\phi (t,{\bf r})$ are Gaussian
independent random functions, $\delta $-correlated in time \cite
{74Kra-a,94SS,95CFKLa}:
\begin{equation}
\langle {\phi (t_1,{\bf r}_1)\phi (t_2,{\bf r}_2)}\rangle =\delta
(t_1-t_2)\chi (r_{12})\ .  \label{a2}
\end{equation}
Here $\chi (r_{12})$ as a function of $r_{12}\equiv |{\bf r}_1-%
{\bf r}_2|$ decays on the scale $L$ and $\chi (0)$ is the production rate of
$\theta ^2$. We consider Prandtl number $Pr=\nu /\kappa $
(viscosity-to-diffusivity ratio) to be large so that $L$ lies in the viscous interval
where the isotropic velocity statistics is described by the pair correlation
function $\langle v_\alpha (t_1,{\bf r}_1)v_\beta (t_2,{\bf r}_2)\rangle $
given by
\begin{equation}
\delta (t_1-t_2)\biggl[V_0\delta _{\alpha \beta }-D\left( \frac 32\delta
^{\alpha \beta }r^2-r^\alpha r^\beta \right) \biggr]  \label{a6}
\end{equation}
for the scales less than the velocity infrared cut-off $L_u$, which is
supposed to be the largest scale of the problem, (\ref{a6}) represents two
first terms of the expansion of the velocity correlation function in $r/L_u$
so that $D\sim V_0/L_u^2$. We presume also the inequality $%
Pe^2=dDL^2/2\kappa \gg 1$ which guarantees that the mean diffusion scale $%
r_d=2\sqrt{\kappa /D}$ is much less than the pumping scale $L$. It follows
from (\ref{a6}) that the correlation functions of the strain field $\sigma
_{\alpha \beta }\equiv \nabla _\beta v_\alpha $ are ${\bf r}$-independent.
That property means that $\sigma _{\alpha \beta }$ can be treated as a
random function of time $t$ only. To exploit that, it is convenient to pass
into the comoving reference frame that is to the frame moving with the
velocity of a Lagrangian particle of the fluid \cite{95CFKLa,96FKLM}:
\begin{equation}
\partial _t\theta +r_j\sigma _{jl}(t)\partial _l\theta -\kappa \Delta \theta
=\phi (t,{\bf r}),  \label{ur}
\end{equation}
Making spatial Fourier transform and introducing
\begin{equation}
\hat{{\cal W}}(t)\equiv T\exp \int_0^t\sigma (\tau )\,d\tau \,,  \label{w}
\end{equation}
one can write the solution of (\ref{ur}) as follows
\begin{eqnarray}
&&\theta _k(t)=\int_0^tdt^{\prime }\phi \left( t-t^{\prime },{\hat{{\cal W}}%
^{-1,T}(t^{\prime }){\bf k}}\right)   \nonumber \\
&&\times \exp \left[ -\kappa k_\mu \int_0^{t^{\prime }}\left[ \hat{{\cal W}}%
^{-1}(\tau )\hat{{\cal W}}^{-1,T}(\tau )\right] _{\mu \nu }d\tau k_\nu
\right] \,.  \label{Lambda}
\end{eqnarray}
Averaging over $\phi $ a simultaneous product of $2n$-th replicas of the
inverse Fourier transform of ${\bf k}\theta _k$, we get the $n$-th moment of
the dissipation field $\epsilon =\kappa \left( \nabla \theta \right) ^2$.
The PDF 
restored from all the moments is
\begin{eqnarray}
&&{\cal P}(\epsilon )=\frac 1{2\pi ^2i}\!\int\limits_{0^{+}-i\infty
}^{0^{+}+i\infty }\!dse^{s\epsilon }\int d{\bf m}e^{-m^2}\bigl\langle e^{-sQ}%
\bigr\rangle_\sigma \,,  \label{PDF} \\
&&Q=\int_0^\infty dt\int d{\bf q}%
\chi _q\left[\frac{{\bf q}\hat{{\cal W}}(t){\bf m}}{2\pi Pe}\right]^2
\exp \left[-\frac{{\bf q}%
\hat{\Lambda}(t){\bf q}}{Pe^2}\right],  \label{QQ} \\
&&\hat{\Lambda}(t)\equiv D\hat{{\cal W}}(t)\int\limits_0^tdt^{\prime }\hat{%
{\cal W}}^{-1}(t^{\prime })\hat{{\cal W}}^{-1,T}(t^{\prime })\hat{{\cal W}}%
^T(t),  \label{H7}
\end{eqnarray}
where ${\bf q}={\bf k}L$ and an extra integration over the auxiliary vector
field ${\bf m}$ takes care of combinatorics and summation over vector
indices. Notice that the direction of time in (\ref{QQ},\ref{H7}) is
opposite to that in (\ref{ur}). The averaging over the traceless $\hat{\sigma%
}=\left(
\begin{array}{cc}
a & b+c \\
b-c & -a
\end{array}
\right) $ is according to ${\cal D}\hat{\sigma}(t)\exp [-\int_0^\infty
dt^{\prime }(a^2+b^2+{c^2}/{2})/2D]$. The whole expression (\ref{QQ}) is
invariant with respect to the global (time independent) rotation ${\bf m}\to
\hat{R}{\bf m}$, $\hat{\sigma}\to \hat{R}\hat{\sigma}\hat{R}^T$, provided
the pumping is statistically isotropic. That allows to get rid of the
angular integration in $d{\bf m}$, counting all the dynamical angles from
the direction of ${\bf m}$.

Let us do the step 2 now. The general tool 
to average a function of $\hat{W}$, particularly (\ref{QQ}), is Kolokolov
transformation replacing $\mbox{T}\exp $ by a regular function of new fields
\cite{90Kol}. We represent $\hat{{\cal W}}=\hat{R}_\vartheta\hat{V}\hat{%
\varphi}$, where $\hat{R}_\vartheta$ is the matrix of rotation by the angle $%
\int_0^tdt^{\prime}\vartheta(t^{\prime})/2$, $\hat{V}$ is diagonal matrix
with the elements $\exp[\pm\int_0^tdt^{\prime}\sigma(t^{\prime})]$ and $\hat{%
\varphi}=\left(
\begin{array}{ll}
1 & \varphi(t) \\
0 & \ 1
\end{array}
\right)$, $\varphi(0)=0$ to provide for $\hat{W}(0)=\hat{1}$. Our transform $%
\{a,b,c\} \rightarrow\{\sigma ,\vartheta ,\varphi \}$ is a hybrid version of
those used in \cite{95BCKLb,95SS} and in Appendix B of \cite{95CFKLa}. The
relation between the old and new variables is obtained by differentiating $%
\hat{R}_\vartheta\hat{V}\hat{\varphi}$ and comparing with $d{\hat W}/dt=\hat{%
\sigma}\hat{W}$:
\begin{eqnarray}
&&a+ib=\sigma \exp \left[i\int\limits_{0}^{t}\!dt^{\prime }\vartheta
\right]+i \frac{\dot{\varphi}}{2}\exp \left[ \int\limits_{0}^{t}dt^{\prime
}\left(2\sigma + i\vartheta\right) \right] ,  \nonumber \\
&&2c=-\vartheta +{\dot{\varphi}}\exp \left[ 2\int_{0}^{t}dt^{\prime }\sigma
\right] .  \label{eu6}
\end{eqnarray}
The Jacobian of the transformation $J=J_{ul}J_{l}$ has a standard product $%
J_{ul}\sim \prod_{t=0}^{T}\exp \left[ 2\int_{0}^{t}dt^{\prime }\sigma \right]
$ and nontrivial Jacobian $J_{l}\sim \exp \left[ \int_{0}^{T}dt\sigma
(t)\right]$ associated with the positive Lyapunov exponent, describing the
exponential stretching of trajectories. The rotation matrix $\hat{R}%
_\vartheta$, which is the same for all the dynamical matrix processes, can
be removed by collective transformation of all the external vectors in the
problem i.e. one may explicitly integrate over ${\cal D}\vartheta$ since we
average $\vartheta$-independent objects in (\ref{PDF}). Averaging some
function of $\hat{{\cal W}}(T)$ is then reduced to averaging the same
function of $\hat{V}(T)\hat{\varphi}$ with respect to the measure ${\cal D}%
\sigma (t){\cal D}\varphi (t)\prod_{t=0}^{T}\exp [2\int_{0}^{t}dt^{\prime
}\sigma ]\exp[-{\cal S}]$, where
\begin{equation}
{\cal S}\equiv\frac{1}{2D} \int\limits_{0}^{T}dt\left\{ \sigma ^{2}-2D\sigma
+\exp \left[ 4\int\limits_{0}^{t}dt\sigma \right] \frac{\dot{\varphi}^{2}}{4}%
\right\} \,.  \label{B8n}
\end{equation}

Finally, we apply the time-separation procedure to capture the term in (\ref
{PDF}) dominant at large ${\rm Pe}$. Two very different time scales are
clearly seen in (\ref{PDF}--\ref{H7}) at least
for $\epsilon\gg\langle\epsilon\rangle/Pe^2$ when the integral over time in (\ref
{QQ}) has to be large. Large value of the integral is achieved on such
realizations of $\hat{\sigma}$ where the integrand grows exponentially due
to ${\cal W}(t)$ until the last exponential factor restricts the grows at
the time of order $D^{-1}\ln {\rm Pe}$. Well before, on a time scale of the
inverse stretching rate $D^{-1}$, the exponentially growing ${\cal W}%
(t^{\prime })$ makes the integral over $dt^{\prime }$ in (\ref{H7})
saturated.

A dynamical picture in $r$-space would be as follows: For the Lagrangian
trajectories giving the main contribution into $\langle\epsilon^n\rangle$,
fluid particles start very close. Diffusion remains dominant until the
particles separate by a distance comparable to the diffusion scale $r_d$.
This phase of the dynamics takes place on times of order $D^{-1}$, notice
that this time is diffusion-independent since $r_d$ is a scale where
diffusion and stretching are comparable. Once the distance between particles
outgrows $r_d$, random multiplicative stretching due to velocity becomes
dominant. Due to a multiplicative nature of the dynamics, the time to go
from $r_d$ to $L$ is proportional to $D^{-1}\ln {\rm Pe}$. Let us now
introduce a separation time $t_{0}$ satisfying $1\ll Dt_{0}\ll \ln [{\rm Pe}%
] $, that time will disappear from the final answer.

The temporal separation makes it possible to substitute $t$ by $t_0$ as an
upper limit in integration over $t^{\prime }$ in (\ref{H7}). Also, an
important manifestation of time separation is a distinctively different
behavior of the $\varphi $ field (responsible for the ``rotation'' in the
pseudospace) at time intervals smaller and larger than the
separation time. The action (\ref{B8n}) shows that for $t>t_0\gg
1/D$ the $\varphi $ field is frozen - no dynamics at all, $\varphi
(t)=\varphi (t_0)$, which has a clear physical meaning since stretching
proceeds along one direction. At the smallest times the $\varphi $-field
dynamics is essential. Indeed, let us denote $\rho (t)=\int_0^t\sigma
(t^{\prime })dt^{\prime }$ and explicitly transform $\hat{\Lambda}(t)$
according to (\ref{eu6}):
\begin{eqnarray}
&&\hat{\Lambda}(t)=\left( \matrix{e^{2\rho}&0\cr0&1}\right)
D\int_0^tdt^{\prime }\exp [{2\rho (t^{\prime })-2\rho (t)}]  \label{Lam} \\
&&\times \left( \matrix{e^{-4\rho(t')}+[\varphi(t)-\varphi(t')]^2&\,%
\varphi(t)-\varphi(t')\cr \varphi(t)-\varphi(t')&1}\right) \left( %
\matrix{e^{2\rho}&\!0\cr0&\!1}\right) .  \nonumber
\end{eqnarray}
For trajectories with predominantly positive $\sigma $ we replace $\varphi
(t)$ by $\varphi (t_0)$ and neglect the integral from $0$ to $t_0$ in $Q$ at
$t\gg t_0$. Since $[\varphi (t)-\varphi (t^{\prime })]^2\exp [2\rho
(t^{\prime })]$ and $\exp [-2\rho (t^{\prime })]$ decrease exponentially as $%
t^{\prime }$ increases, then the integral over $dt^{\prime }$ saturates,
which allows for time separation too. Finally, non-diagonal and lower
diagonal elements in $\hat{\Lambda}$ are exponentially small. For time
separation to be complete, the integral $\rho (t)$ at $t\gg t_0$ has to be
counted from $t_0$ i.e. we replace $\rho (t)\to \rho (t)+\rho (t_0)$, then
in the dominant order in P\'{e}clet we get
\begin{eqnarray}
&&Q\!=\!{\frac{m^2\beta }{(2\pi Pe)^2}}\!\int\limits_{t_0}^\infty \!dt\!\int
d{\bf q}\chi _qq_1^2e^{2\rho }\exp \left( -q_1^2e^{2\rho }\frac{\mu \beta }{%
Pe^2}\right) ,  \label{SS3} \\
&&\mu \equiv D\int_0^{t_0}dt^{\prime }e^{2\rho (t^{\prime })}\{e^{-4\rho
(t^{\prime })}+[\varphi (t)-\varphi (t^{\prime })]^2\},  \label{SS2}
\end{eqnarray}
where we denoted $\beta =\exp [2\rho (t_0)]$. Next, one needs to average $%
\exp \left[ -sQ\right] $ over the short-time ${\cal D}\sigma _{<}{\cal D}%
\varphi $ and a long-time ${\cal D}\sigma _{>}$ separately. The
corresponding weights of averaging with respect to $\hat{\sigma}_{>}$ and $%
\hat{\sigma}_{<}$ are completely decomposed: ${\cal S}={\cal S}_{<}+{\cal S}%
_{>}$. The great advantage of (\ref{SS3}) is that, in the long-time
averaging, both $\beta $ and $\mu $ are just external parameters, depending
neither on time $t$ nor on $\sigma _{>}$. Once the average over $\sigma _{>}$
is performed, we are left with a function of $\mu $. The final result for $%
{\cal P}(\epsilon )$ will be then obtained by averaging over $\sigma _{<}$
and $\varphi $.

We start from averaging $\exp [-sQ]$ over $\sigma _{>}$ which reduces to the
following auxiliary quantum mechanics
\begin{eqnarray}
&&{\cal P}_s^{>}\left( {\bf m},\mu \right) =\left[ e^{-(T-t_0)/2}\Phi
(T-t_0;\rho )\right] _{\rho =0},  \label{Ls1} \\
&&\left( \partial _t-\hat{H}_{>}\right) \Phi =0,\quad \hat{H}_{>}\equiv -%
\frac 12\partial _\rho ^2  \label{Ls2} \\
&&+\frac{sm^2\beta }{\left( 2\pi \right) ^2\mbox{Pe}^2}\int d{\bf q}\chi
_qq_1^2e^{2\rho }\exp \left[ -q_1^2e^{2\rho (t)}\frac{\mu \beta }{Pe^2}%
\right] .  \nonumber
\end{eqnarray}
Here $sQ$, with $Q$ from (\ref{SS3}), plays the role of potential, the
quadratic in $\sigma $ term of (\ref{B8n}) gives the derivative part of the
Hamiltonian, while the linear in $\sigma $ part of (\ref{B8n}) gives the
initial condition at $t=0$ for the ``wave function'': $\Phi (0;\rho )=e^\rho
$. 
Since the ``potential'' part of the quantum mechanics vanishes at $\rho \to
\infty $ and $\Phi (0;\rho )$ does not, one obtains the asymptotic behavior
of $\Phi $ at $t\to \infty $
\begin{eqnarray}
&&\Phi (t;\rho )\to e^{t/2}y\mbox{Pe}\Phi _{*}(y),\ {\cal P}_s^{>}=\Phi
_{*}((\mbox{Pe}\mu \beta )^{-1})  \label{Phi} \\
&&\left( -\frac 1{2y^3}\partial _yy^3\partial _y+\frac{sm^2}\mu U\left(
y\right) \right) \Phi _{*}(y)=0,  \label{deq} \\
&&U(x)\equiv \frac 1{\left( 2\pi \right) ^2}\int d{\bf k}\chi _kk_1^2\exp
\left[ -k_1^2x^2\right] ,  \label{pot}
\end{eqnarray}
where the new variable $y=e^\rho (\mbox{Pe}\mu \beta )^{-1}$ has been
introduced. $\Phi _{*}\to 1$ at $y\to \infty $ and $\Phi _{*}y$ should
vanish at $y\to 0$. The generating function is a function of a single
argument $sm^2/\mu $, we denote ${\cal P}_s^{>}=C(2sm^2/\mu )$. The
potential $U(x)$ is everywhere positive, it has to turn into a constant at $%
x\to 0$ and behave as $x^{-3}$ at large $x$. Under such a general assumption
on the pumping function $\chi _q$, one can show that the only singularities
of $C(z)$ are poles on the negative semiaxis at a finite distance from zero
(see \cite{97CFKb} for the details). The asymptotic condition on $\Phi
_{*}(y)$ gives the normalization $C\left( 0\right) =1$ and at $z\rightarrow
\infty $ $\ln [1/C(z)]\sim \sqrt{z}$. For example, in the particular case
\begin{equation}
U_c(x)=%
{1,\hspace{0.5cm}x<1, \atopwithdelims\{. 1/x^3,\hspace{0.5cm}x>1.}
\label{Phi3}
\end{equation}
one finds $2C(z)=z\left[ I_2^{\prime }(2\sqrt{z})I_1(\sqrt{z})+I_2(2\sqrt{z}%
)I_1^{\prime }(\sqrt{z})\right] ^{-1}$ and $C(z)\to ({z^{3/2}}/{4\pi })\exp
\left[ -3\sqrt{z}\right] $ at ${z\to +\infty }$.

Now we have to average ${\cal P}_s^{>}=C(2sm^2/\mu )$ over $\sigma _{<}$ and
$\varphi $, what is equivalent to averaging over the random variable $\mu $
(responsible for the fluctuations of the diffusion scale due to strain
variations): $\langle e^{-sQ}\rangle _\sigma =\langle C(2sm^2/\mu )\rangle
_\mu $. A convenient auxiliary object to calculate is the generating
function ${\cal P}_\lambda =\left\langle \exp \left[ -\lambda \mu \right]
\right\rangle $, which can be found from yet another quantum mechanics:
\begin{eqnarray}
&&{\cal P}_\lambda =\int_{-\infty }^\infty d\varphi \Psi \left( \rho
=0,\varphi \right) ,\ \ \left( \hat{H}_{<}+\frac 12\right) \Psi =0,
\label{small} \\
&&\left[ \int_{-\infty }^\infty d\varphi \varphi ^n\Psi e^{-\rho }\right]
_{\rho \rightarrow +\infty }\rightarrow \delta _{n0},\quad \Psi \left|
_{\rho \rightarrow -\infty }\rightarrow 0\right. .  \nonumber \\
&&\hat{H}_{<}\equiv -2e^{-4\rho }\partial _\varphi ^2-\frac 12\partial _\rho
^2+\lambda e^{-2\rho }\left( 1+\varphi ^2e^{4\rho }\right) ,  \nonumber
\end{eqnarray}
where the static equation (\ref{small}) follows at $t_0\gg 1$ from the
respective dynamical one similarly to the way (\ref{Phi}) follows from (\ref
{Ls1}). Exact solution of (\ref{small}), $\Psi \pi \sqrt{1+\varphi ^2e^{4\rho }}=%
\sqrt{2\lambda }e^{2\rho }K_1\left( e^{-\rho }\sqrt{2\lambda (1+\varphi
^2e^{4\rho })}\right) $ gives 
\begin{equation}
{\cal P}(\mu )=\left( 2\pi \mu ^3\right) ^{-1/2}\exp \left( -1/[2\mu
]\right) .  \label{PDFmu}
\end{equation}
Note that ${\mu}^{-1/2}$ (which can be interpreted as inverse local diffusion scale)
has exactly Gaussian PDF, that seems to be a consequence of strain Gaussianity.
Integration of $C(2sm^2/\mu )$ with ${\cal P}(\mu )$ gives the final answer:
\begin{equation}
{\cal P}(\epsilon )=\frac 1{2\pi \sqrt{\epsilon }}\int\limits_{-\infty
}^\infty \!dsC\left( is\right) \int\limits_0^\infty \!dx\exp \left( isx^2-%
\frac{\sqrt{\epsilon }}x\right) .\hspace{0.1cm}  \label{pdfe}
\end{equation}
That formula is our main result. It expresses dissipation PDF in terms of
the function $C$ determined by (\ref{deq}) with the potential (\ref{pot})
given by the pumping. For any pumping, all is necessary to get ${\cal P}%
(\epsilon )$ is to solve an ODE of the second order. In particular, one can
find $C(z)$ as a series in $z$ iterating $\Psi $ in (\ref{deq}) over $U$.
That expresses dissipation moments directly via the pumping
\begin{eqnarray}
&&\langle \epsilon ^n\rangle =2^n\left[ (2n-1)!!\right] ^2n!\times
\label{epsn} \\
&&\int\limits_0^\infty \frac{dy_1}{y_1^3}\int\limits_0^{y_1}dy_2y_2^3U(y_2)%
\cdots \int\limits_{y_{2n-2}}^\infty \frac{dy_{2n-1}}{y_{2n-1}^3}%
\int\limits_0^{y_{2n}}dy_{2n}y_{2n}^3U(y_{2n}).  \nonumber
\end{eqnarray}
One gets, particularly, the conservation law identity $\langle \epsilon
\rangle =-2C^{\prime \prime }(0)=\chi [0]/2$ as a consistency check of our
procedure. The form of the PDF at $\epsilon \simeq \langle \epsilon \rangle $
is pumping-dependent while the asymptotics at small and large $\epsilon $
have universal forms.

At small $\epsilon $ (and infinite $Pe$) the two first terms have the form
${\cal P}(\epsilon )\to {A\epsilon^{-1/2}+B\ln \epsilon }$ 
so that gradient's PDF ${\cal P}(\nabla \theta )\to A-B|\nabla \theta |\ln
|\nabla \theta |^{-1}$ tends to a constant (equilibrium equipartition),
notice nonanalyticity of the first non-equilibrium
correction. The asymptotics is sensitive to the fast angular
dynamics.

Since the only singularities of $C(iz)$ are poles on the imaginary axis then
large-$\epsilon $ asymptotics is determined by the pole nearest to the
origin in the upper half-plane: $z=i\tau ^{*},\tau ^{*}\sim 1$. In
particular, $\tau_*\approx2.7$ for (\ref{Phi3}). The $dx$-integration in (%
\ref{pdfe}) is done by a saddle-point method:
\begin{equation}
{\cal P}(\epsilon )\sim {\epsilon }^{-1/2}\exp \left( - 3(2\tau ^{*}\epsilon
)^{1/3}/2\right) .  \label{bolshas}
\end{equation}
Stretched-exponential tail is natural 
for steady PDF of the gradients \cite{94SS} contrary to lognormal unsteady
distribution which takes place without diffusion \cite{74Kra-a}. The value $%
1/3$ for the stretched exponent may be explained as follows: Since the local
value of $\nabla\theta$ is proportional to the scalar fluctuation at
the diffusion scale $\delta\theta$ (which has exponential PDF tail 
\cite{94SS,95CFKLa,97BGK}) times an inverse local diffusion scale 
[which is Gaussian according to (\ref{PDFmu})] then $%
\langle\epsilon^n \rangle\sim
\kappa^n\langle(\delta\theta)^2\rangle^nn^{2n}r_d^{-2n}n^n$ which corresponds
to (\ref{bolshas}). Note that the asymptotics (\ref{bolshas})
is determined by the dynamics of stretching (not of
rotations), thus it is likely to take place in any dimensions. The
data of numerics done directly for Kraichnan model \cite{BDFL} are
satisfactory fitted by $1/3$ stretched exponent within the window expanding
when $Pe$ grows. The exponent $1/3$ agrees also with the values $0.3\div 0.36
$ given by numerics \cite{94HS} and $%
0.37$ by experimental data on air turbulence \cite{95OR}. Moreover, our
exponent derived formally at $Pr\gg 1$ agree with the data of \cite
{94HS,95OR} obtained at $Pr\simeq 1$ as well. This is surprising
because ${\cal P}(\epsilon )$ is independent of $Pe$ at large $Pr$ while at $%
Pr\lesssim1$ the statistics of $\epsilon$ is markedly different since $%
\langle \epsilon ^n\rangle \propto Pe^{\mu _n}$ where $\mu _n$ are anomalous
exponents of the scalar \cite{96CF}. That may be, however, that the PDF tail
is still given by (\ref{bolshas}) for any $Pr$ with $\tau_*$ and preexponent
factor in (\ref{bolshas}) generally depending on both $Pr$ and $Pe$, this
will be the subject of future studies. 
To conclude, the agreement of our result with a variety of data
suggests that Kraichnan model is a proper tool for recovering exponents in
the theory of turbulent mixing.

We thank E. Balkovsky, V. Lebedev, B. Shraiman, and M. Vergassola for
interesting discussions. This work was partially supported by a R. H. Dicke
fellowship (MC), the Israel Science Foundation (GF) and Russian Fund of
Fundamental Researches under grant 97-02-18483 (IK).

\end{multicols}

\end{document}